\documentclass[12pt] {article}
\usepackage{graphicx,graphics}
\usepackage{epsfig}

\newfont{\jnp}{cmss10}
  
      \def\beq{\begin{eqnarray}}
  \def\eeq{\end{eqnarray}}    
  \def\be{\begin{equation}}  \def\ee{\end{equation}}
\begin{document}
{\bf \Large Properties of $B_c$ meson}\\

{\bf \large Ajay Kumar Rai and  P C Vinodkumar}\\
Department of Physics, Sardar Patel University,
Vallabh Vidyanagar, Gujarat-388120, India.\\
{\bf keywords}:-Potential, Decay Constant, Weak decay.\\
{\bf pacs No}:-14.40.Nd\\
Email:-raiajayk@rediffmail.com \\

{\bf Abstract}\\
The mass spectrum of $c \bar b$ meson is investigated with an
effective quark-antiquark potential of the form $\frac{-\alpha
_c}{r}$ + $A \ r^{\nu}$ with $\nu$ varying from 0.5 to 2.0. The S
and P-wave masses, pseudoscalar decay constant, weak decay partial
widths in spectator model and the lifetime of $B_c$ meson are
computed. The properties calculated here are found to be in good
agreement with other theoretical and experimental values at
potential index, $\nu=1$.

\section{Introduction}
The discovery of the $B_c$ meson by the Collider Detector at
Fermilab (CDF Collaboration) \cite{CDFcollabration} in $p \bar p$
collision at $\sqrt{s}$ = 1.8 TeV has demonstrated the possibility
of the experimental study of the charm-beauty system and has created
considerable interest in its spectroscopy
\cite{Buchmuller,Eichten,ALVhady,EFGEbert,Latticedav,lpfulcher,Gershtein}.
It is the only meson in the heavy flavour sector with different
charge and flavours, due to which its decay properties are expected
to be different from that of flavour neutral mesons. Though there
exist results on charmed hadrons suggesting the importance of
relativistic effects, however, studies based on nonrelativistic
models also provide results close to the experimental values
\cite{CDFcollabration,Buchmuller,Eichten,ALVhady,Altarelli,Hagiwara,dshwang}.

\section{Nonrelativistic treatment for heavy quarks}
 For a heavy-heavy quark bound system such as $c \bar
b$ we treat both the quarks c and $\bar b$ nonrealtivistically.
The Hamiltonian for this case is given by \cite{Ajayk}
 \begin{equation}
\label{eq:nlham} H= M + \frac{p^2}{2M_1} + V(r)\end{equation}
where
 \begin{equation} \label{eq:mm1} M=m_c + m_{\bar b} , \  \ \  \ and \   \  \ \
M_1=\frac{m_c \ m_{\bar b}}{m_c + m_{\bar b}}\end{equation}
 $m_c$ and $m_{\bar b}$ are the mass parameters of charmed quark
 and bottom quark respectively, p is the relative momentum of each quark and V(r)
is the quark antiquark potential. We consider here a general power
potential with coulomb term of the form,
 \begin{equation}
\label{eq:potver} V(r)=\frac{-\alpha_c}{r} + A r^\nu
\end{equation}
where $\alpha _c = \frac{4}{3} \alpha_s $,  $ \alpha _s $ being the
strong running coupling constant, $A$ is a model potential parameter
and $\nu $ is a general power corresponding to the confining part of
the potential. In the present case, we study the system by varying
$\nu$ from 0.5 to 2.0 and the parameter $A=0.19 \ GeV^{\nu+1}$, $
m_b=4.66 \ GeV$, $ m_c=1.31 \ GeV$, are taken to
be the same as used in the study of light-heavy flavour mesons \cite{Ajayk}.\\
Within the Ritz variational scheme we assume a trial radial wave
function $R(r,\ \mu)$, and compute the expectation value of the
Hamiltonian given by  Eqn.(\ref{eq:nlham}) ($<H>=E(\mu ,\nu ) \ $)
with the potential defined by Eqn.(\ref{eq:potver}). For the ground
state we get,
  \begin{equation}
\label{eq:toteg}E(\mu, \nu )=M+\frac{1}{8} \
\frac{\mu^2}{M_1}+\frac{1}{2}\left(- \mu \ \alpha_c + A \ \frac{
\Gamma(\nu+3)} {\mu^{\nu}}\right) \end{equation} The  trial wave
function  is assumed to be of the form
\begin{equation} \label{eq:wavfun} R_{nl}(r) = \left(\frac{\mu^3
(n-l-1)! }{2n(n+l)!}\right)^{1/2} \ \  (\mu \ r)^l \ \ e^{- \mu r
/2} \ \ L^{2l+1}_{n-l-1}(\mu r)\end{equation}
 Here,  $\mu$ is the variational parameter and $L^{2l+1}_{n-l-1}(\mu r)$
 is Laguerre polynomial.
For a chosen value of $\nu$, the variational parameter, $\mu $ is
determined for each state using the virial theorem
\begin{equation}
\label{eq:virial}
 \left<\frac{P^2}{2 M_1}\right>=\frac{1}{2}\left<\frac{r
 d V}{dr}\right> \end{equation}
As the interaction potential assumed here, does not contain the spin
dependent part,  Eqn.(\ref{eq:toteg}) gives the spin average masses
of the system in  terms of the power index $\nu$. The spin average
mass for the ground state is computed and are listed in
Table-1 for the values of $\nu$ from 0.5 to 2.0.  \\
\begin{table}[hbt]
 \label{tbspnag}
\caption{ The variational parameter ${\bar \mu }$, wave function
at the origin ($|R(0)|$) S-wave and P-wave (Spin average) masses
of $c \bar {b}$ meson.} \hskip4pc\vbox{\columnwidth=26pc
\begin{tabular}{llllllll}
Mesonic&State& $\nu$ & ${ \bar \mu }$ &
$|R(0)|$&\multicolumn{1}{c}{$E(\bar \mu ) $}
 &EFG \cite{EFGEbert}&ZVR \cite{JZeng}\\
Systems&&&$GeV$&$GeV^{3/2}$&(GeV)&(GeV)&(GeV)\\
\hline
&&0.5&1.1920&0.920&6.230&&\\
&1S&1.0&1.6020&1.434&6.367&6.317&6.320\\
&&1.5&1.9580&2.358&6.509&$6.387^*$&\\
&&2.0&2.2760&2.428&6.657&&\\
\hline
&&0.5&1.0910&0.254&6.419&&\\
&1P&1.0&1.6960&0.765&6.738&6.736&6.740\\
&&1.5&2.2302&1.516&7.073&&\\
&&2.0&2.7040&2.454&7.414&&\\
\hline
&&0.5&1.1250&0.844&6.457&&\\
$c \bar {b}$&2S&1.0&1.7950&1.701&6.846&6.869&6.887\\
&&1.5&2.3965&2.623&7.269&&\\
&&2.0&2.9330&3.552&7.699&&\\
\hline
&&0.5&1.1360&0.459&6.566&&\\
&2P&1.0&1.9440&1.756&7.136&7.142&7.150\\
&&1.5&2.7035&4.006&7.799&&\\
&&2.0&3.3985&7.097&8.505&&\\
\hline
&&0.5&1.1528&0.875&6.586&&\\
&3S&1.0&1.993&1.989&7.201&7.224&7.270\\
&&1.5&2.7874&3.291&7.923&&\\
&&2.0&3.5174&4.665&8.696&&\\
\hline
\end{tabular}}
 $^{*}$AlV\cite{ALVhady}, $\alpha_s =0.255 $, $ m_b=4.66 \ GeV$, $ m_c=1.31 \
GeV$, $A=0.19 \ GeV^{\nu+1} $.
\end{table}
\noindent  For the S-wave and P-wave mass calculations we consider
the spin-spin and spin-orbit interactions as \cite{Gershtein}
\begin{equation} \label{spindependent}
V_{S_c \ \cdot \ S_b}(r)= \frac{8}{9} \frac{ \alpha_s}{ m_c m_b} \
\vec S_c \ \cdot  \vec S_b \ 4 \pi \delta(r); \ \ \ \ V_{L\ \cdot
\ S}(r)= \frac{4 \ \alpha_s}{3 \ m_c m_b} \frac{\vec L \ \cdot
\vec S}{r^3} \end{equation}
The computed masses are compared with
other theoretical predictions of Eichten and Quigg \cite{Eichten},
A L Hady\cite{ALVhady}, D Ebert et al. \cite{EFGEbert}, C T H
Davies et al. \cite{Latticedav} and Gershtein et al.
\cite{Gershtein} in Table-2. Our predicted mass for $B_c(1 ^1S_0)$
is in good accord with experimental result of 6.40$ \pm$ 0.39
(stat) $\pm$ 0.13(syst) GeV/$C^2$ \cite{CDFcollabration} and the
masses obtained for the 2S, 3S, 1P, 2P states are comparable with
other theoretical predictions.
\section{Decay properties of $B^+_c$ meson:}
\noindent The Decay properties of $B^+_c$ ($\bar b c$) meson is of
interest as it decays only through weak interactions
\cite{CDFcollabration,ALVhady,EFGEbert}. This is due to the fact
that its ground state energy lies below the (BD) threshold and has
non vanishing flavour. This eliminates the uncertainties
encountered due to strong decays and provides a clear decay width
and lifetime for $B^+_c$ meson,  which helps to fix more precise
value of the weak decay parameters such as the CKM mixing matrix
elements ($V_{cb},\ V_{cs}$) and the leptonic decay constant
($f_p$). Adopting the spectator model for the charm-beauty system
\cite{ALVhady}, the total decay width of $B^+_c$ meson can be
approximated  as the sum of the widths of $\bar b$-quark decay
keeping c-quark as spectator, the c-quark decay with $\bar
b$-quark as spectator, and the annihilation channel $B^+_{c}
\rightarrow l^+\nu_l (c \bar s, \ u \bar s)$, $l=e,\mu,\tau$ with
no interference assumed between them. \\

\begin{table}\label{tb:spectrm}
\caption{$B_c$ meson mass spectrum (in GeV) with $\nu$=1}
\begin{center}
\begin{tabular}{llllll}
 \hline
 $n^{2S+1}L_J$& $Our$
&ALV\cite{ALVhady}&EQ\cite{Eichten}&EFG\cite{EFGEbert}
&Lattice\cite{Latticedav}\\
\hline
1$^1S_0$&6.349&6.356&6.264&6.270&6.280$\pm 30 \pm 190$\\
1$^3S_1$&6.373&6.397&6.337&6.332&6.321$\pm 20$\\
1$^3P_0$&6.715&6.673&6.700&6.699&6.727$\pm 30$\\
1$^3P_1$&6.726&-&6.730&6.734&6.743$\pm 30$\\
$1$$^1P_1$&6.738&-&6.736&6.749&6.765$\pm 30$\\
1$^3P_2$&6.749&6.751&6.747&6.762&6.783$\pm 30$\\
2$^1S_0$&6.821&6.888&6.856&6.835&6.960$\pm 80$\\
2$^3S_1$&6.855&6.910&6.899&7.072&6.990$\pm 80$\\
2$^3P_0$&7.102&-&7.108&7.091&-\\
2$^3P_1$&7.119&-&7.135&7.126&-\\
2$^1P_1$&7.136&-&7.142&7.145&-\\
2 $^3P_2$&7.153&-&7.153&7.156&-\\
3$^1S_0$&7.175&-&7.244&7.193&-\\
3$^3S_1$&7.210&-&7.280&7.235&-\\
\hline
\end{tabular}
\end{center}
\end{table}
\noindent Accordingly, the total width is written as
\cite{ALVhady}
\begin{equation} \label{eq:totalwidth}
 \Gamma(B_c \rightarrow X)= \Gamma(b \rightarrow X)+ \Gamma(c \rightarrow X)
 +\Gamma(Anni)\end{equation}
Neglecting the quark binding effects, we obtain for the b and c
inclusive widths in the spectator approximation \cite{ALVhady},
 \begin{equation}
 \Gamma(b \rightarrow X)= \frac{ 9 \ G^2_F \left|V_{cb}\right|^2
 m^5_b}{192 \pi^3}= 8.75  \times 10^{-4} eV  \end{equation}
  \begin{equation} \label{restol}
 \Gamma(c \rightarrow X)= \frac{ 5 \ G^2_F \left|V_{cs}\right|^2
 m^5_c}{192 \pi^3} = 4.19 \times 10^{-4} eV  \end{equation}
 Here we have used $\left|V_{cs}\right|=0.975$,
$\left|V_{cb}\right|=0.044$ as the upper bound provided by particle
data group \cite{Hagiwara}, and the value of $m_b$,
$m_c$ as used in our mass predictions.\\
In the nonrelativistic limit, the pseudoscalar constant $f_P$ and
the ground state wave function  at the origin $R(0)$ are related
and is given by the Van Royen Weisskopf \cite{Eichten} formula
including the color factor, as
  \begin{equation} \label{decay con.f_p} f_{B_c}=\sqrt{\frac{3}{\pi
M_{B_c}}}R_{1s}(0).\end{equation}
 The $f_{B_c}$ values obtained here are 361 MeV, 556 MeV, 757 MeV and  929 MeV for
 $\nu$ = 0.5, 1.0, 1.5 and 2.0 respectively.\\
 Now, the width of the annihilation channel is computed using the expression
 given by \cite{ALVhady}
 \begin{equation} \label{ttlannrest}
  \Gamma(Anni)= \frac{G^2_F}{8 \pi} \left|V_{bc}\right|^2 f^2_{B_C} M_{bc}
\sum_i m^2_i\left(1- \frac{m^2_i}{M^2_{B_c}}\right)^2 \ . \ C_i,
 \end{equation}
\[ =\ 0.923 \ \times \ 10^{-4}  \ eV\]
 Where $C_i=1$ for the $\tau \nu_{\tau}$ channel and $C_i=3\left|V_{cs}\right|^2$
 for $c \bar s$, and $m_i$ is the mass of the heaviest fermions.
 Our result for $f_{B_c}$ and $M_{B_c}$ obtained with the
 potential parameter $\nu=1$ are used in Eqn(\ref{ttlannrest}).\\
\noindent Adding all the three contributions according to
Eqn(\ref{eq:totalwidth}) yield the total width
$\Gamma(total)=13.863 \times 10^{-4} eV$ and the lifetime of
$B^+_c$ meson as $0.47 ps$, which is in good agreement with the
measured value of $\tau=0.46^{+0. 18} _ { -0.16} ps $
\cite{CDFcollabration}.

\begin{table} \label{ta:pdcm}
 \caption{Comparison of the lifetime of $B_c$ meson (in ps) in
different models.} \vspace{0.5cm}
\begin{center}
 \begin{tabular}{cccccc}
\hline Our&Expt\cite{CDFcollabration}& ALV\cite{ALVhady}& GKLT
\cite{Gershtein} &VVK \cite{VVKiselev}&
SG\cite{GodfreyS}\\
 \hline
0.47&$\tau=0.46^{+0.18}_{-0.16}$&0.47&0.55$\pm 0.15$&0.50&0.75\\
\hline
\end{tabular}
\end{center}
\end{table}
\section{Conclusion}
\noindent Based on a simple nonrelativistic potential scheme within
variational approach we have been able to predict the S-wave, P-wave
masses and lifetime of $B^+_c$ meson successfully. Mass spectrum and
pseudoscalar decay constants $f_{B_c}$ are computed for the
potential index ($\nu$) from 0.5 to 2. Our predictions of the masses
and $f_{B_c}$ values are found to be in accordance with other
theoretical predictions for  $\nu\ = \ 1$. It is found that
$f_{B_c}$ and $M_{B_c}$ increases as the potential parameter $\nu $ increases.\\

\noindent The model parameters such as the charm and beauty quark
masses used in our calculations and the pseudoscalar decay constants
$f_{B_c}$ obtained here for $\nu\simeq 1$ are found to be
appropriate in the calculation of the decay widths. We get about $63
\%$ as the branching fractions of b-quark decay, about $30 \%$ as
that of c-quark decay and about $7 \%$ in the annihilation channel.
However the CKM mixing matrix elements $V_{cb}$ and $V_{cs}$ used as
free parameters in all the theoretical calculations compared here
are different but within the range given in particle data group
\cite{Hagiwara}. The lifetime of $B^+_{c}$ predicted by the present
calculation is found to be in good agreement  with the experimental
values as well as that by the Bethe Salpeter method(ALV)(See
Table-3). The predicted values from relativized model (SG) is found
to be far from the experimental values as well as other theoretical
models.\\

\noindent In conclusion, a simple nonrelativistic variational
method with potential $ -\frac{\alpha_c}{r}+A r^{\nu} $ employed
in the present study is found to be quite successful in predicting
various properties of $B^+_c$ meson. The method can be useful to
study various hadronic and radiative transitions of the
charm-beauty system.


\begin{thebibliography}{99}
\bibitem{CDFcollabration}CDF Collabration, Phys. Rev. D {\bf 58}, 112004
(1998).
\bibitem{Buchmuller}Buchmuller W and  Tye S H H, Phy. Rev. D {\bf
24} 132 (1981).
\bibitem{Eichten}Eichten E J and  Quigg , Phys. Rev. D {\bf 49},
 5845 (1994).
\bibitem{ALVhady} A. Adb El-Hady, M.A.K. Lodhi and J.P. Vary, Phys. Rev D.{\bf
  59} 094001 (1999).
\bibitem{EFGEbert} D. Ebert, R. N. Faustov and V. O. Galkin, Phys Rev. D {\bf
67} 014027 (2003), [hep-ph/0210381]
\bibitem{Latticedav}C T H Davies, K Hornbostel, G P Lepage, A J Lidsey, J Shigemitsu
 and J h sloan, Phys, Lett. B {382} 131 (1996), [arXiv:hep-lat/9602020]
 (The error estimates are taken from \cite{lpfulcher}, \cite{GodfreyS} and reference their in)
 \bibitem{lpfulcher} L P Fulcher,Phys. Rev D.{\bf 60} 074006 (1999).
\bibitem{Gershtein} Gershtein S S, Kiselev V V, Likhoded A K and Tkabladze A V,
Phys. Rev. D {\bf 51}, 3613 (1995).
\bibitem{Altarelli} Altarelli G, Cabbibbo N, Corbo G, Maiani Land Martinelli
G, Nucl. Phys. { \bf B 208.} 365 (1982).
\bibitem{Hagiwara} Hagiwara K et. al, Particle Data Group,  Phys. Rev. D.{\bf
66}, 010001-1 (2002)
 \bibitem{dshwang} D S Hwang, C S Kim and Wuk Namgung, Phys. Rev. D.{\bf
53}, 4951 (1996)
\bibitem{Ajayk} Ajay K Rai, Parmar R H and Vinodkumar P C, J. Phys G.
{\bf 28}  2275 (2002); Ajay Kumar Rai, Pandya J N and Vinodkumar P
C, J. Phys G. {\bf 31}  1453 (2005).
\bibitem{JZeng} J Zeng, J W Van Orden and W Roberts, Phys. Rev. D {\bf 52}, 5229(1995).
\bibitem{VVKiselev}V V Kiselev, arXiv:hep-ph/0308214.
\bibitem{GodfreyS} Stephen Godfrey, arXiv:hep-ph/0406228(2004).
\end{thebibliography}
\end{document}